\documentclass[onecolumn,10pt]{revtex4}

\usepackage{graphicx}
\usepackage{dcolumn}
\usepackage{bm}
\usepackage{color}
\usepackage{amssymb,amsmath}

\input epsf
\begin{document}

\title{\Large{The effects of the pole dark energy on gravitational waves}}

\author{\bf Homa Shababi$^1$\footnote{h.shababi@scu.edu.cn}, Sayani Maity$^2$\footnote{sayani.maity88@gmail.com}, Prabir Rudra$^3$\footnote{prudra.math@gmail.com}, Ujjal Debnath$^4$\footnote{ujjaldebnath@gmail.com}}
\affiliation{$^1$Center for Theoretical Physics, College of Physical Science and Technology, Sichuan University, Chengdu 610065, People’s Republic of China.\\
$^2$Department of Mathematics, Sister Nivedita University,
DG-1/2, Action Area 1, New Town, Kolkata-700 156, India\\
$^3$Department of Mathematics, Asutosh College, Kolkata-700 026, India.\\
$^4$Department of Mathematics, Indian Institute of Engineering Science and Technology,
Shibpur, Howrah-711 103, India.\\}

\date{\today}

\begin{abstract}
In this paper, we have studied the effects of pole dark energy on the evolution of gravitational waves. The background evolution of gravitational waves in a flat FRW universe is considered and its dynamics are studied in the presence of pole dark energy. Two different potential functions are considered for the study. Using the field equations, we formulated the perturbed equations governing the evolution of gravitational waves with respect to redshift $z$ within the background of the FRW Universe. Subsequently, we delved into the characteristics of gravitational waves for the pole dark energy model and reached interesting results.
\end{abstract}

\maketitle

\section{Introduction}
Gravitational waves, a fascinating consequence of Einstein's general theory of relativity, have been a subject of intrigue for physicists and astronomers alike. In the realm of cosmology and astrophysics, gravitational waves play a pivotal role. These ripples in spacetime carry valuable information about their amplitude, which can be extracted from measurements of cosmic microwave temperature anisotropy and polarization \cite{Crittenden1,Crittenden2}. The idea that disturbances in the gravitational field can propagate as waves appears to be intuitive and mirrors the behavior of other types of waves we observe in nature. Einstein demonstrated that gravitational radiation, in the form of gravitational waves, is a natural outcome of his theory, thereby solidifying the connection between gravitational waves and the fundamental principles of general relativity. In the limit of small deviations from Euclidean space-time (or Minkowski space), Einstein's field equations yield a linear wave equation with plane wave solutions. These solutions describe transverse metric perturbations of Minkowski space that travel at the speed of light, exhibiting characteristics analogous to those of electromagnetic waves. While there are several similarities between gravitational and electromagnetic waves, it is essential to note that this comparison has its limitations and should be approached with caution. In the early Universe, gravitons underwent decoupling, leading to the separation of gravitational waves from matter. Consequently, these gravitational waves play a crucial role in constraining and distinguishing cosmological parameters across various cosmological models. Notably, primordial gravitational waves originated from vacuum fluctuations. In a hypothetical scenario during the early Universe, where a seamless transition occurs between an early de Sitter-like phase and a radiation-dominated era, the matter content can be described by a cosmological model that aims to unify the dark sectors of the Universe—namely, dark energy and dark matter. Among all the conceivable models of dark energy in cosmology, gravitational waves can provide valuable insights into epochs when there were fluctuations in the underlying cosmic dynamics. Up to know, a lot of investigations have been done about gravitational waves \cite{Buonanno97,Gasperini1,Fabris1,Infante,Riazuelo,Fabris2,Santos,Lopez,Zhang,Debnath1,Debnath2}.

According to recent observations, there is strong evidence that our Universe is currently undergoing accelerated expansion. Clues from various sources, including type Ia supernovae, the Cosmic Microwave Background (CMB), large-scale structure (LSS), and WMAP observations, all point to this cosmic acceleration \cite{Supernova,Perlmutter,Spergel,Briddle,Liddle,Spergel2,Komatsu}. The driving force behind this expansion is an enigmatic substance known as Dark Energy. Dark Energy possesses positive energy density and sufficient negative pressure, which violates the strong energy condition, that is $\omega = {p}/{\rho} < -1/3$. Numerous theoretical models have been put forth by researchers to elucidate the enigmatic nature of dark energy \cite{Copland, Li2}. Despite these efforts, the true essence of dark energy remains elusive. Among the simplest contenders is the cosmological constant $\Lambda$ with $\omega=-1$, originally introduced by Einstein to account for the Universe’s static behavior. Another straightforward candidate is the quintessence, characterized by a spatially homogeneous scalar field $\phi$ that gradually descends a potential as $\phi\rightarrow\infty$ \cite{Peebles, Caldwell}. The scientific literature also features other dark energy candidates, including tachyon, Chaplygin gas, holographic dark energy, dilaton, k-essence, DBI-essence, hessence, and new agegraphic dark energy, all of which play essential roles in driving the Universe’s acceleration \cite{Sen1,Sen2,Gasperini,Amendola,Armendariz1,Armendariz2,Wei1,Wei2,Gumjudpai,Martin,Ogawa,Kamenshchik,Li,Hsu,Wei3,Wei4}.

The presence of a pole in the kinetic term has proven highly effective in the study of inflation within theoretical frameworks. These pole kinetic terms exhibit quantum stability and attractor properties, making them valuable tools. While a scalar field with a pole in its kinetic term is commonly employed for studying cosmological inflation, it can also serve as dark energy, leading to what is known as the pole dark energy model \cite{Feng}. When poles are applied to dark energy theories, they reveal an intriguing connection between thawing and freezing models, as well as the potential for enhanced plateaus characterized by “superattractor-like” behavior. Notably, even simple models incorporating pole dark energy can yield an equation-of-state evolution with $\omega(z)<-0.9$, a feat that would typically be challenging for other potential forms. Furthermore, the kinetic term pole provides an interesting perspective in relation to the swampland criteria for observationally viable dark energy models \cite{Linder}.

Herein, our primary objective is to explore the characteristics of gravitational waves within the context of pole dark energy model \cite{Linder} in a Friedmann-Robertson-Walker (FRW) background of the Universe. To this end, we investigate the impact of pole dark energy on the dynamics of gravitational waves. We examine the evolution of gravitational waves in a flat FRW universe while accounting for the influence of pole dark energy. Our study explores two distinct potential functions including a power law, and an exponential potential form. The work is structured as follows: Background equations of gravitational waves in a flat FRW universe are studied in section II. The dynamics of the gravitational waves in the background of pole dark energy are explored in section III. Finally the paper ends with a brief conclusion in section IV.

\section{Background Equations of Gravitational Waves in a Flat FRW Universe}

For standard and flat FRW metric
\begin{equation}\label{2.1}
ds^2=-dt^2+a^2(t)\left(dx^2+dy^2+dz^2\right),
\end{equation}
we can obtain the Einstein’s field equations as (with $8\pi G=c=1$) \cite{Debnath1}
\begin{equation}\label{2.2}
\frac{\dot a^2}{a^2}=\frac{1}{3}\left(\rho_m+\rho_D\right),
\end{equation}
\begin{equation}\label{2.3}
\frac{2\ddot{a}}{a}+\frac{\dot a^2}{a^2}=-\left(p_m+p_D\right),
\end{equation}
where $a(t)$ is the scale factor, $\rho_m$ is the energy density of dark matter, $\rho_D$ is the energy density of dark energy, and $p_m,  p_D$ are the pressures of dark matter and dark energy, respectively.

Given the assumption of separate conservation for dark matter and dark energy, the continuity equations for them are given by
\begin{equation}\label{2.4}
\dot\rho_m+\frac{3\dot a}{a}\left(\rho_m+p_m\right)=0,
\end{equation}
and
\begin{equation}\label{2.5}
\dot\rho_D+\frac{3\dot a}{a}\left(\rho_D+p_D\right)=0.
\end{equation}
Considering dark matter satisfies the equation of state $p_m=\omega_m\rho_m$, where $\omega_m$ is the constant equation of state parameter, we obtain
\begin{equation}\label{2.6}
\rho_m=\rho_{m_0}a^{-3\left(1+\omega_m\right)}=\rho_{m_0}\left(1+z\right)^{3\left(1+\omega_m\right)},
\end{equation}
where $\rho_{m_0}$ is the present value of the dark matter density and $Z=\frac{a_0}{a}-1$ is the redshift.

The governing equations of the evolution of gravitational waves for flat universe are as follows:
\begin{equation}\label{2.7}
    \ddot{\eta}(t)-\frac{\dot{a}}{a}\dot{\eta}
(t)+\left(\frac{\xi^2}{a^2}-2\frac{\ddot{a}}{a} \right)\eta(t)=0,
\end{equation}
which can be written as:
\begin{equation}\label{2.8}
    \eta^{''}(z)-\left(\frac{\ddot{a}}{\dot{a}^2}-\frac{2}{a} \right)a^2 \eta^{'}(z)
    +\frac{a^4}{\ddot{\eta}^2}\left(\frac{\xi^2}{a^2}-2\frac{\ddot{a}}{a} \right)\eta(z)=0,
\end{equation}
where $'\equiv \frac{d}{dz}$.

Next we express the equations (\ref{2.2}) and (\ref{2.3}) as follows:
\begin{equation}\label{2.9}
\frac{\dot a^2}{a^2}=H_0^2 X(z)~~~ \mbox{and}~~~\frac{2\ddot{a}}{a}=-H_0^2 Y(z),
\end{equation}
where
\begin{equation}\label{2.10}
    X(z)=\frac{1}{3H_0^2}\left(\rho_m+\rho_D \right),
\end{equation}
and
\begin{equation}\label{2.11}
    Y(z)=\frac{1}{3H_0^2}\left[\rho_m \left(1+3\omega_m \right)+\rho_D\left(1+3\omega_c\right)\right],
\end{equation}
where $\omega_m=\frac{p_m}{\rho_m}, \omega_c=\frac{p_D}{\rho_D}$ are the equation of state parameter of dark matter and dark energy respectively, and $H_0$ is the present value of the Hubble parameter $H=\frac{\dot a}{a}$.
Exploiting equations (\ref{2.9}), (\ref{2.10}) and (\ref{2.11}), equation (\ref{2.8}) reads:
\begin{equation}\label{2.12}
    \eta^{''}(z)+\frac{1}{2(1+z)}\left( \frac{Y(z)}{X(z)}+4\right) \eta^{'}(z)
    +\frac{4}{Y^2(z)H_0^4}\left(\xi^2+\frac{H_0^2 Y(z)}{(1+z)^2}\right)\eta(z)=0.
\end{equation}
 The characteristic of the gravitational waves for pole dark energy model in flat FRW Universe is going to be investigated in the following sections.

\section{Gravitational Waves for Pole Dark Energy Model}
Lagrangian for pole dark energy is defined as \cite{Linder}
\begin{equation}\label{3.1}
L_{\sigma}=-\frac{1}{2}\frac{\kappa}{\sigma^{p}}(\partial\sigma)^{2}-V(\sigma),
\end{equation}

which the pole can be positioned at $\sigma=0$ without any loss of generality. It possesses a residue of $k$ and an order of $p$. Poles can emerge in theories due to nonminimal coupling with the gravitational sector, geometric characteristics of the K\"{a}hler manifold in supergravity, or as an indication of soft symmetry breaking (as discussed in references \cite{Broy,Terada}). In our treatment, we approach it from a phenomenological perspective.
Using the transformations

\begin{equation}\label{3.2}
\phi=\frac{2\sqrt{k}}{|2-p|}\sigma^{\frac{2-p}{2}}, ~~~~~~\sigma=\left(\frac{|2-p|}{2\sqrt{k}}\right)^{\frac{2}{2-p}}\phi^{\frac{2}{2-p}},
\end{equation}
we get canonical form the Lagrangian as
\begin{equation}\label{3.3}
L_{\sigma}=-\frac{1}{2}\left(\partial\phi\right)^2-V(\sigma).
\end{equation}

Using Eq. \eqref{3.3}, density and pressure in the standard form for FRW universe respectively express as
\begin{equation}\label{3.4}
\rho_{\phi}=\frac{\dot\phi^{2}}{2}+V(\phi),~~~~p_{\phi}=\frac{\dot\phi^{2}}{2}-V(\phi),
\end{equation}

where dot shows the derivative with respect to time.\\

For the case $p=2$, we have
\begin{equation}\label{3.5}
\phi=\pm\sqrt{k}\ln\sigma, ~~~~~~\sigma=e^{\pm\frac{\phi}{\sqrt{k}}}.
\end{equation}

Now, from Eq. \eqref{3.4}, we obtain
\begin{equation}\label{3.6}
\dot \rho_\phi=\ddot{\phi}\dot \phi+V'(\phi)\dot \phi=\dot \phi\left(\ddot{\phi}+V'(\phi)\right).
\end{equation}
The conservation equation of the canonical form of pole dark energy reads
\begin{equation}\label{3.7}
\dot\rho_\phi+\frac{3\dot a}{a}\left(\rho_\phi+p_\phi\right)=0.
\end{equation}

Substituting Eqs. \eqref{3.4} and \eqref{3.5} in Eq. \eqref{3.7}, we find
\begin{equation}\label{3.8}
\dot\phi\left(\ddot{\phi}+V'(\phi)\right)+3\frac{\dot a}{a}{\dot\phi}^2=0.
\end{equation}
Next we take $V(\phi)=\frac{1}{2}V_0 \dot{\phi}^2$ and insert in equation \eqref{3.8} that yields
\begin{equation}\label{3.9}
\dot{\phi}=\phi_0 a^{-\frac{3}{1+v_0}}=\phi_0 (1+z)^{\frac{3}{1+v_0}} ~~\mbox{and}~~V(\phi)=\frac{1}{2}V_0\phi_0^2 (1+z)^{\frac{6}{1+v_0}}.
\end{equation}
Inserting the expression of $V(\phi)$ from equation \eqref{3.9} in \eqref{3.4} we get the energy density and pressure of canonical form of pole dark energy as follows:
\begin{equation}\label{3.10}
\rho_\phi=\frac{1}{2}(1+V_0)\phi_0^2 (1+z)^{\frac{6}{1+v_0}}~~\mbox{and}~~p_\phi=\frac{1}{2}(1-V_0)\phi_0^2 (1+z)^{\frac{6}{1+v_0}},
\end{equation}
which can be expressed as
\begin{equation}\label{3.11}
\rho_\phi=\rho_{\phi 0} (1+z)^{\frac{6}{1+v_0}}~~\mbox{and}~~p_\phi=p_{\phi 0} (1+z)^{\frac{6}{1+v_0}},
\end{equation}
where $\rho_{\phi 0} =\frac{1}{2}(1+V_0)\phi_0^2$ and $p_{\phi 0}=\frac{1}{2}(1-V_0)\phi_0^2$ are the present values of the energy density and pressure of pole dark energy respectively.
Now, we define the dimensionless density parameters as
\begin{equation}\label{3.12}
\Omega_{m0}=\frac{\rho_{m0}}{3H_0^2},~~\mbox{and}~~\Omega_{c0}=\frac{\rho_{D0}}{3H_0^2},
\end{equation}
where $\rho_{m0}$ and $\rho_{D0}$ are the present values of the energy density of the matter and dark energy respectively.\\

Using equations \eqref{2.6}, \eqref{3.11} and \eqref{3.12} in the field equation \eqref{2.2} we get
\begin{equation}\label{3.13}
H(z)=H_0\left[ \Omega_{m0}(1+z)^{3(1+\omega_m)}+\Omega_{c0}(1+z)^{\frac{6}{1+v_0}}\right]^{\frac{1}{2}}.
\end{equation}
At present epoch from \eqref{3.13} can be written as
\begin{equation}\label{3.14}
\Omega_{m0}+\Omega_{c0}=1.
\end{equation}
For canonical form of pole dark energy model the evolution equation \eqref{2.12} of gravitational wave is the function of redshift $z$ with observed cosmological parameters $H_0, \omega_m, \Omega_{m0}, \Omega{c0}$, where $X(z)$, $Y(z)$ can be  computed from \eqref{2.10} and \eqref{2.11} as follows
\begin{equation}\label{3.15}
X(z)=\Omega_{m0}(1+z)^{3(1+\omega_m)}+\Omega_{c0}(1+z)^{\frac{6}{1+v_0}},
\end{equation}
and
\begin{equation}\label{3.16}
Y(z)=\Omega_{m0}(1+z)^{3(1+\omega_m)}+2\Omega_{c0}\left(\frac{2-v_0}{1-v_0}\right)(1+z)^{\frac{6}{1+v_0}}.
\end{equation}
Since it is very difficult to solve the differential equation (\ref{2.12}) by inserting the expressions of $X(z)$ and $Y(z)$ from (\ref{3.15}) and (\ref{3.16}) for Pole Dark Energy model, we do not generate the analytical solution of $\eta(z)$. Here we plot the wave function $\eta(z)$ against various ranges of redshift $z$ in figure 1, figure 2 and figure 3 for $H_0=67$, $\omega_m=0.001$, $v_0=1.5$, $\Omega_{m0}=0.26$, $\Omega_{c0}=0.74$, $\xi=10^8$ (black dotted line) and $\xi=2\times 10^9$(blue line).

\begin{figure}[hbt!]
\centering
\includegraphics{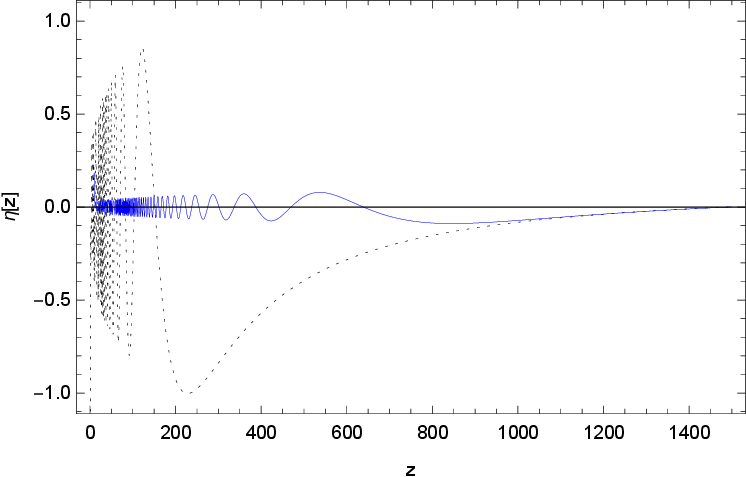}
\centering \caption{Plot of $\eta(z)$ against redshift $0 \leq z \leq 1500$ for past universe with $\xi=10^8$ (black dotted line) and $\xi=2\times10^9$ (blue line).}
\label{Fig1a}
\end{figure}

\begin{figure}[hbt!]
\centering
\includegraphics{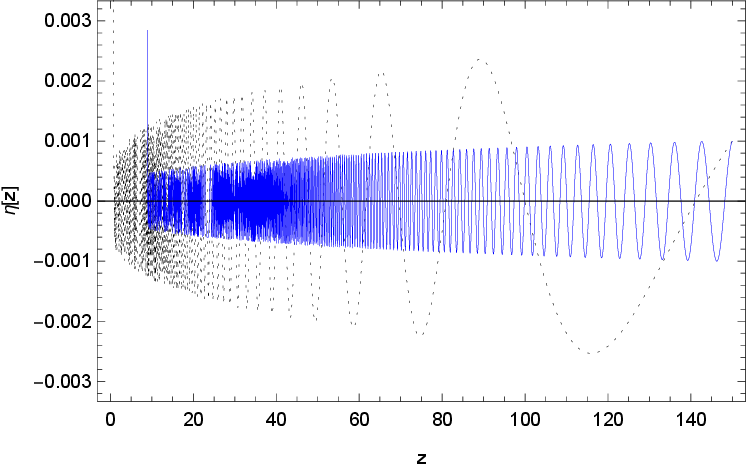}
\centering \caption{Plot of $\eta(z)$ against redshift $0 \leq z \leq 200$ with  $\xi=10^8$ (black dotted line) and $\xi=2\times10^9$ (blue line) to give a more magnified picture around the present universe.}
\label{Fig1a}
\end{figure}

\begin{figure}[hbt!]
\centering
\includegraphics{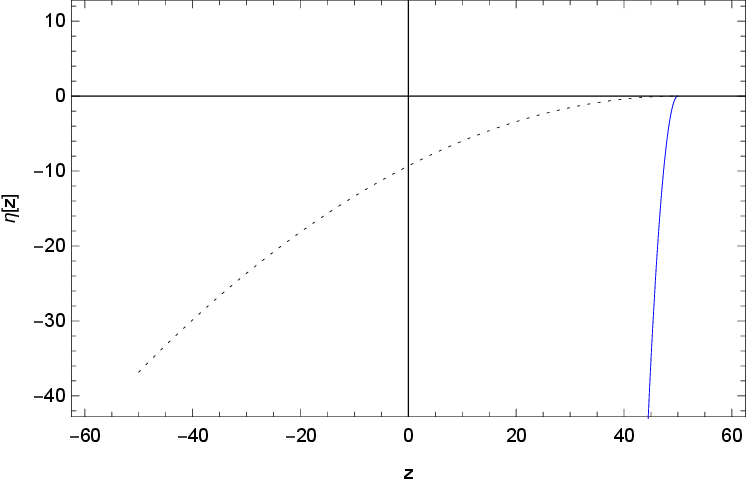}
\centering \caption{Plot of $\eta(z)$ against redshift $-50 \leq z \leq 50$ that includes future universe with $\xi=10^8$ (black dotted line) and $\xi=2\times10^9$ (blue line).}
\label{Fig1a}
\end{figure}

\subsection{Power law potential}
A power law potential is transformed into a power law potential as
\begin{equation}\label{3.1.1}
V\sim\sigma^n\Rightarrow V\sim\phi^{\frac{2n}{2-p}},
\end{equation}
which for $p=1$
\begin{equation}\label{3.1.2}
V\sim\sigma^n\Rightarrow V\sim\phi^{2n},
\end{equation}
and for $p=4$
\begin{equation}\label{3.1.3}
V\sim\sigma^n\Rightarrow V\sim\phi^{-n},
\end{equation}
Now the derivation of Eq. \eqref{3.1.1} leads
\begin{equation}\label{3.1.4}
V'(\phi)=\frac{2n}{2-p}\phi^\frac{2n-2+p}{2-p}.
\end{equation}
Substituting \eqref{3.1.4} in \eqref{3.8} we gain
\begin{equation}\label{18}
\dot\phi\left(\ddot{\phi}+\frac{2n}{2-p}\phi^\frac{2n-2+p}{2-p}\right)+3\frac{\dot a}{a}{\dot\phi}^2=0.
\end{equation}
Considering $\dot\phi\neq0$, we have
\begin{equation}\label{19}
\frac{\ddot{\phi}}{\dot\phi}+\frac{2n}{2-p}\frac{\phi^{\frac{2n-2+p}{2-p}}}{\dot\phi}=-\frac{3\dot a}{a}.
\end{equation}
Since this equation is not directly solvable, we need to make some transformations. We consider the field in terms of the scale factor in a power law form as
\begin{equation}\label{20}
\phi(a)=\phi_0 a^m,
\end{equation}
where $\phi_0$ and $m$ are constants.
So we get,
\begin{equation}\label{21}
\dot\phi(a)=\phi_0 m a^{m-1}\dot a,
\end{equation}
and
\begin{equation}\label{22}
\ddot{\phi}(a)=\phi_0 m \left(a^{m-1}\ddot a+(m-1)a^{m-2}\dot a^2\right).
\end{equation}
Now putting all these equations in Eq. \eqref{19}, we obtain a differential equation for the scale factor $a(t)$,
\begin{equation}\label{23}
\ddot{a}(t)+\frac{(m+2)\dot a^2}{a(t)}+\frac{2n}{m(2-p)}{\phi_0}^{\frac{2(n+p-2)}{2-p}}{a(t)}^{\frac{2mn+2mp-4m-p+2}{2-p}}=0.
\end{equation}

The solution of Eq. \eqref{23} is obtained as,
\begin{equation}\label{****}
a(t)=InvFunc\left[\frac{\sqrt{m(6+mn-3p)}{}_2F_1\left[\frac{1}{2},-\frac{\left(3+m\right)\left(-2+p\right)}{2\left(6+mn-3p\right)},1-\frac{\left(3+m\right)\left(-2+p\right)}{2\left(6+mn-3p\right)},\frac{2n\phi_{0}^{-\frac{2\left(-2+n+p\right)}{-2+p}}}{m\left(6+mn-3p\right)C_{1}}\sqrt{\frac{1-2n\phi_{0}^{-\frac{2\left(-2+n+p\right)}{-2+p}}}{m\left(6+mn-3p\right)C_{1}}}\right]}
{\left(3+m\right)\sqrt{\phi_{0}^{\frac{-2\left(n+p\right)}{-2+p}}\left(-2n\phi_{0}^{\frac{4}{-2+p}}+m\left(6+mn-3p\right)\phi_{0}^{\frac{2\left(n+p\right)}{-2+p}}C_{1}\right)}}\right]\left[t+C_{2}\right],
\end{equation}
where 'InvFunc' represents InverseFunction and $_2F_1$ represents hypergeometric function. Moreover $C_{1}$, $C_{2}$ are arbitrary constants.

Putting the above expression for $a(t)$ in Eq. \eqref{20}, we get
\begin{equation}\label{*****}
\phi(t)=\phi_0\left(InvFunc\left[\frac{\sqrt{m(6+mn-3p)}{}_2F_1\left[\frac{1}{2},-\frac{\left(3+m\right)\left(-2+p\right)}{2\left(6+mn-3p\right)},1-\frac{\left(3+m\right)\left(-2+p\right)}{2\left(6+mn-3p\right)},\frac{2n\phi_{0}^{-\frac{2\left(-2+n+p\right)}{-2+p}}}{m\left(6+mn-3p\right)C_{1}}\sqrt{\frac{1-2n\phi_{0}^{-\frac{2\left(-2+n+p\right)}{-2+p}}}{m\left(6+mn-3p\right)C_{1}}}\right]}
{\left(3+m\right)\sqrt{\phi_{0}^{\frac{-2\left(n+p\right)}{-2+p}}\left(-2n\phi_{0}^{\frac{4}{-2+p}}+m\left(6+mn-3p\right)\phi_{0}^{\frac{2\left(n+p\right)}{-2+p}}C_{1}\right)}}\right]\left[t+C_{2}\right]\right)^m.
\end{equation}

\subsection{Exponential potential}
For an exponential potential, we have
\begin{equation}\label{*}
V\sim e^{-\lambda\sigma}\Rightarrow V\sim e^{\frac{-\lambda\sqrt{k}}{\phi}},
\end{equation}
\begin{equation}\label{**}
V'(\phi)=\frac{\lambda\sqrt{k}}{\phi^2}e^{\frac{-\lambda\sqrt{k}}{\phi}}.
\end{equation}
Now, we put Eq. \eqref{*} into Eq. \eqref{3.7}
\begin{equation}\label{***}
\dot\phi\left(\ddot{\phi}+\frac{\lambda\sqrt{k}}{\phi^2}e^{\frac{-\lambda\sqrt{k}}{\phi}}\right)+3\frac{\dot a}{a}{\dot\phi}^2=0.
\end{equation}
With $\dot\phi\neq0$, we have
\begin{equation}\label{****}
\frac{\ddot{\phi}}{\dot\phi}+\lambda\sqrt{k}\frac{\phi^{-2}e^{\frac{-\lambda\sqrt{k}}{\phi}}}{\phi}=-\frac{3\dot a}{a}.
\end{equation}

Using the transformation $\phi(a)=\phi_0 a^m$, we obtain
\begin{equation}\label{V2}
\ddot{a}(t)+\frac{(m+2)\dot a(t)^2}{a(t)}+\frac{\lambda\sqrt{k}}{m}{\phi_0}^{-3}e^{\frac{-\lambda\sqrt{k}}{\phi_0 a(t)^m}}{a(t)}^{-3m+1}=0,
\end{equation}
which leads to the following solution
\begin{equation}\label{SV2}
a(t)=InvFunc\left[\frac{{}_2F_1\left[\frac{1}{2},\frac{3+m}{6-m},\frac{9}{6-m},-\frac{2e^{-\sqrt{k}}\lambda\phi_{0} a^{-m}\sqrt{k}\lambda}{\left(-6+m\right)m\phi_{0}^{3} C_{3}}\right]\sqrt{1+\frac{2e^{-\sqrt{k}}\lambda\phi_{0} a^{-m}\sqrt{k}\lambda}{\left(-6+m\right)m\phi_{0}^{3} C_{3}}}}
{\left(3+m\right)\sqrt{\frac{2e^{-\sqrt{k}}\lambda\phi_{0} a^{-m}\sqrt{k}\lambda}{\left(-6+m\right)m\phi_{0}^{3}}+C_3}}\right]\left[t+C_4\right],
\end{equation}

where $C_3$ and $C_{4}$ are the constant of integration.
Now, substituting Eq. \eqref{SV2} in \eqref{20}, we find
\begin{equation}\label{SV3}
\phi(t)=\phi_0\left(InvFunc\left[\frac{{}_2F_1\left[\frac{1}{2},\frac{3+m}{6-m},\frac{9}{6-m},-\frac{2e^{-\sqrt{k}}\lambda\phi_{0}\sqrt{k}\lambda}{\left(-6+m\right)m\phi_{0}^{3} C_{3}}\right]\sqrt{1+\frac{2e^{-\sqrt{k}}\lambda\phi_{0}\sqrt{k}\lambda}{\left(-6+m\right)m\phi_{0}^{3} C_{3}}}}
{\left(3+m\right)\sqrt{\frac{2e^{-\sqrt{k}}\lambda\phi_{0}\sqrt{k}\lambda}{\left(-6+m\right)m\phi_{0}^{3}}+C_3}}\right]\left[t+C_4\right]\right)^m.
\end{equation}

\section{Conclusion}
In our study, we investigated the impact of pole dark energy on the evolution of gravitational waves. Our motivation for this study stems from the success of incorporating poles in the kinetic term when studying inflation within theoretical frameworks. These pole kinetic terms demonstrate quantum stability and attractor properties, rendering them valuable tools. In this direction, firstly we explored the flat Friedmann-Robertson-Walker (FRW) model of the Universe, taking into account both dark matter and dark energy. By analyzing separate conservation equations for dark matter and dark energy, we considered energy densities for both components. Then, we explored the properties of gravitational waves in the context of the pole dark energy model within a flat FRW Universe. Through specific transformations, we derived the canonical Lagrangian for pole dark energy. By utilizing this Lagrangian, we obtained expressions for the density and pressure in the standard form applicable to the FRW universe.
Using the field equations, we formulated the perturbed equations governing the evolution of gravitational waves with respect to redshift $z$ within the background of the FRW Universe. Subsequently, we delved into the characteristics of gravitational waves for the pole dark energy model. Due to the complexity of the differential equations associated with gravitational waves in this model, we employed graphical analysis to obtain wave curves across various redshift ranges (as depicted in Figs. 1, 2, and 3). In all the plots, we utilized the observed values of the parameters $H_0=67$, $\omega_m=0.001$, $v_0=1.5$, $\Omega_{m0}=0.26$, $\Omega_{c0}=0.74$, $\xi=10^8$ (black dotted line) and $\xi=2\times 10^9$(blue line). As we see in all figures, the amplitudes increase over time, as $z\rightarrow0$. Moreover, for more investigations, we examined two distinct potential functions in our study and obtained the results.

\section*{Acknowledgments}

PR acknowledges the Inter-University Centre for Astronomy and Astrophysics (IUCAA), Pune, India for granting visiting associateship.

\end{document}